\documentclass[conference]{IEEEtran}
\IEEEoverridecommandlockouts
\usepackage{lscape}
\usepackage{rotating}
\usepackage{epstopdf}
\usepackage{etoolbox}
\usepackage{nomencl}
\usepackage{supertabular}
\usepackage{amsfonts}
\usepackage[dvips]{epsfig}
\usepackage{latexsym}
\usepackage{amssymb}
\usepackage{amsmath}
\usepackage{verbatim}
\usepackage{setspace}
\usepackage[ruled,linesnumbered,boxed,commentsnumbered]{algorithm2e}
\usepackage{bm}
\usepackage[colorlinks=true, linkcolor=black, citecolor=blue, urlcolor=blue]{hyperref}       % hyperlinks
\usepackage{url}            % simple URL typesetting
\usepackage{lipsum}	

\usepackage{cite}
\usepackage{amsmath,amssymb,amsfonts}
\usepackage{algorithmic}
\usepackage{graphicx}
\usepackage{picinpar}
\usepackage{flushend}
\usepackage[latin1]{inputenc}
\usepackage{colortbl}
\usepackage{soul}
\usepackage{pifont}
\usepackage{alltt}
\usepackage{enumerate}
\usepackage{breakurl}
\usepackage{subfigure}
\usepackage{epstopdf}
\usepackage{pbox}
\usepackage{fancyhdr}

\usepackage{stfloats}
\usepackage{url}

\usepackage[ruled,linesnumbered,boxed,commentsnumbered]{algorithm2e}
\usepackage[dvips]{epsfig}
\usepackage{lineno}
\modulolinenumbers[5]
\usepackage{multirow}
\usepackage{comment}

\usepackage{xcolor}
\usepackage{underscore}
\usepackage{optidef}
\def\BibTeX{{\rm B\kern-.05em{\sc i\kern-.025em b}\kern-.08em
    T\kern-.1667em\lower.7ex\hbox{E}\kern-.125emX}}
\usepackage[figure]{hypcap}
\hypersetup{hypertex=true,
            colorlinks=true,
            linkcolor=blue,
            anchorcolor=blue,
            citecolor=blue}

\begin{document}

\title{Network-Aware Value Stacking of Community Battery via Asynchronous Distributed Optimization\\
\thanks{This work is supported in part by the Australian Research Council (ARC) Discovery Early Career Researcher Award (DECRA) under Grant DE230100046.}
}

\author{\IEEEauthorblockN{Canchen Jiang\textsuperscript{1}, Hao Wang\textsuperscript{1,2*}}
\IEEEauthorblockA{\textsuperscript{1}Department of Data Science and AI, Faculty of IT, Monash University, Australia \\
\textsuperscript{2}Monash Energy Institute, Monash University, Australia\\
}
\thanks{*Corresponding author: Hao Wang (hao.wang2@monash.edu).}
}

\maketitle

\begin{abstract}
Community battery systems have been widely deployed to provide services to the grid. Unlike a single battery storage system in the community, coordinating multiple community batteries can further unlock their value, enhancing the viability of community battery solutions. However, the centralized control of community batteries relies on the full information of the system, which is less practical and may even lead to privacy leakage. In this paper, we formulate a value-stacking optimization problem for community batteries to interact with local solar, buildings, and the grid, within distribution network constraints. We then propose a distributed algorithm using asynchronous distributed alternate direction method of multipliers (ADMM) to solve the problem. Our algorithm is robust to communication latency between community batteries and the grid while preserving the operational privacy. The simulation results demonstrate the convergence of our proposed asynchronous distributed ADMM algorithm. We also evaluate the electricity cost and the contribution of each value stream in the value-stacking problem for community batteries using real-world data. 
\end{abstract}

\section{Introduction}
As part of the commitment to achieving net-zero emissions, the the widespread uptake of distributed energy resources (DERs) holds immense potential for decarbonizing the energy system. The community battery stands as one of the promising technologies, finding various applications across the energy system, including peak shaving, reserve support, and managing variability caused by renewable resources \cite{wang2020improving,10252189}.

Recently, there has been a strong interest in deploying community batteries to provide multiple services in the energy system, establishing the value stacking of community battery systems. For example, Tripathy et al. \cite{tripathy2018establishing} assessed the cumulative value of a battery storage system by delivering multiple grid services, such as peak shaving, frequency regulation, and reserve support. Seward et al. \cite{seward2022revenue} utilized the battery storage system to provide auxiliary services to frequency response to generate multiple revenues. However, these studies only considers a single battery system, while it is expected that the coordination of multiple batteries can lead to more benefits but also demands effective methodologies. 

Recognizing the need to maximize the value of the community batteries, their coordination has been studied in recent literature. For example, Chen et al. \cite{chen2023multi} developed an operation strategy for grid-connected battery systems considering energy-sharing among communities. Berg et al. \cite{berg2023quantifying} explored the operation of multiple batteries within an energy community, examining how shared battery usage contributes to providing services for the distribution network. However, centralized solution methods for optimizing community battery coordination may not be practical due to the privacy concerns associated with accessing information from the distribution network and communities. Therefore, distributed solution methods have emerged as a promising solution for coordinating community batteries while preserving their operational privacy. 

The alternate direction method of multipliers (ADMM) has been applied as a decomposition method in designing distributed algorithms for power system operations \cite{olivella2020centralised}. Existing studies have explored privacy-preserving solution methods for residential communities using ADMM algorithms. For example, a privacy-preserving ADMM algorithm was developed in \cite{yang2021privacy} for a blockchain-based transactive energy management system for smart homes. Umer et al. \cite{umer2021novel} developed an energy trading distributed ADMM algorithm to maximize the system social welfare while making the trading scheme more privacy-preserving. However, most of the distributed algorithms rely on a critical condition, in which all the participants need to synchronize information exchange to solve the optimization problem. This assumption may not be practical in real-world scenarios due to unreliable communications or latency, thus hindering the applications of distributed algorithms in practice.

In this paper, we are motivated to develop a value-stacking optimization problem to maximize the value of community batteries through multiple value streams, while considering the distribution network constraints. We propose an asynchronous distributed ADMM algorithm, which allows asynchronous information exchange between communities and the grid operator when solving the network-aware value-stacking problem in an iterative fashion. We use real-world data to evaluate our developed value-stacking optimization and asynchronous distributed algorithm. 

The contributions of this paper are as follows.
\begin{itemize}
    \item We develop a value-stacking optimization problem to enable multiple value streams for community batteries, including battery-to-building, battery-to-grid, and energy trading, taking into account the network constraints for a more reliable assessment. 
    \item We design a distributed optimization algorithm based on the asynchronous ADMM to solve the community battery value-stacking optimization problem. Our algorithm is demonstrated to be convergent, accommodating asynchronous information exchange between communities and the grid operator.
    \item We assess the marginal contribution of each value stream within the context of Australia's National Electricity Market (NEM). We also compare the performance under two different tariffs, e.g., time-of-use (TOU) pricing and two-part tariff (TPT) pricing. 
\end{itemize}

\section{System Model and Problem Formulation}\label{System Model}
We consider an energy system, comprising residential buildings, co-located community batteries, and photovoltaics (PV) on a distribution network. Community battery can interact with the grid, buildings, and other batteries on the network. We denote $\mathcal{I}=\left\{1,\dots, I\right\}$ as the set of nodes on the distribution network, indexed by $i$. Residential communities with buildings and co-located community batteries are connected to some nodes. We denote the operational horizon as $\mathcal{H}=\left\{1,\dots, H\right\}$ with evenly-sliced time slots. The size of each time slot is denoted as $\Delta t$. In our simulations, we set $H=24$ for a one-day operational horizon, and thus $\Delta t = 1$.
 
\subsection{Residential Community Model}
\subsubsection{Community Battery Model}
In community $i$, there is a community battery co-located with buildings. We denote $p_{\text{Batc}}^{i,t}$ and $p_{\text{Batd}}^{i,t}$ as battery charge and dischage power in $t$-th time slot. The stored energy in the battery $b_{\text{Bats}}^{i,t}$ evolves with charging and discharge. The community battery operation satisfies the following constraints
\begin{align}\label{eq1}
    & b_{\text{Bats}}^{i,t} = b_{\text{Bats}}^{i,t-1} + \mu p_{\text{Batc}}^{i,t}\Delta t - \frac{1}{\eta}p_{\text{Batd}}^{i,t}\Delta t, \\
    & \underline{P}_{\text{Bats}}^{i} \leq b_{\text{Bats}}^{i,t} \leq \overline{P}_{\text{Bats}}^{i}, \\
    & 0 \leq p_{\text{Batc}}^{i,t} \leq \overline{P}_{\text{Batc}}^{i,t} (1-x^{i,t}),~t\in \mathcal{H}, \label{eq3}\\
    & 0 \leq p_{\text{Batd}}^{i,t} \leq \overline{P}_{\text{Batd}}^{i,t} x^{i,t},~t\in \mathcal{H}, \label{eq4}
\end{align}
where $\mu\in [0,1]$ and $\eta\in (0,1]$ represent the charging and discharging efficiencies of community battery $i$; $\overline{P}_{\text{Bats}}^{i}$ and $\underline{P}_{\text{Bats}}^{i}$ represent the upper and lower bounds for $b_{\text{Bats}}^{i,t}$; $\overline{P}_{\text{Batc}}^{i,t}$ and $\overline{P}_{\text{Batd}}^{i,t}$ are the maximum limits for charging and discharging, respectively. Since the community battery cannot charge and discharge simultaneously, we introduce a binary variable $x^{i,t}\in \{ 0, 1\}$ to limit its operation.

Community battery charge/discharge will incur degradation, and we consider the following degradation cost:
$
   \mathbf{C}_{\text{battery}}^{i} = \alpha_b \sum \nolimits_{t\in \mathcal{H}} \left( (p_{\text{Batd}}^{i,t}\Delta t)^2+(p_{\text{Batc}}^{i,t}\Delta t)^2 \right),
$
where $\alpha_b$ is the amortized cost coefficient of the battery operation.

\subsubsection{Residential Building Load Model}
The building load comprises various components. In our work, we focus on the Heating, ventilation, and air conditioning (HVAC) load, denoted as $p_{\text{AC}}^{i,t}$, as it often constitutes a significant portion of the building load and is considered flexible. The rest of the load is referred to as inflexible load $ P_{\text{Bd,load}}^{i,t}$. 
The HVAC system controls the indoor temperature $\text{T}_{\text{InB}}^{i,t}$ based on the building $i$'s desired setting, such as a set-point $\text{T}_{\text{InB}}^{i,Pref}$. The building indoor temperature $\text{T}_{\text{InB}}^{i,t}$ evolves with the change of outdoor temperature $\text{T}_{\text{OutB}}^{i,t}$ and HVAC power \cite{cui2019peer} as follows
\begin{equation} \label{eq5}
     \text{T}_{\text{InB}}^{i,t} = \text{T}_{\text{InB}}^{i,t-1} - \frac{1}{C^{i}R^{i}}(\text{T}_{\text{InB}}^{i,t-1} - \text{T}_{\text{OutB}}^{i,t} + \mu^{i} R^{i} p_{\text{AC}}^{i,t} \Delta t),
\end{equation}
where $\mu^{i}$ is the working mode of the HVAC, where positive and negative value represents cooling and heating respectively; $C_{i}$ and $R_{i}$ are the working parameter of the HVAC system. Note that the HVAC load $p_{\text{AC}}^{i,t}$ is bounded as 
\begin{equation}
     \underline{P}_{\text{AC}}^{i,t} \leq p_{\text{AC}}^{i,t} \leq \overline{P}_{\text{AC}}^{i,t},\label{eq6}
\end{equation} 
where $\underline{P}_{\text{AC}}^{i,t}$ and $\overline{P}_{\text{AC}}^{i,t}$ represent the lower bound and upper bound of HVAC load in time slot $t$ for building $i$. 

Any deviation of the indoor temperature from the set-point can cause discomfort for residents, and thus we model the discomfort cost as 
$
     \mathbf{C}_{\text{AC}}^{i} = \beta^{i} \sum_{t \in \mathcal{H}} (\text{T}_{\text{InB}}^{i,t} - \text{T}_{\text{InB}}^{i,Pref})^{2}, 
$
where $\beta^{i}$ is the sensitive coefficient. In addition, the indoor temperature cannot deviate too much and should be kept within a range, denoted as $[\underline{T}^{i,t}, \overline{T}^{i,t}]$ throughout all time slots. Hence, we model the indoor temperature limits as
\begin{equation} \label{eq7}
     \underline{T}^{i,t} \leq \text{T}_{\text{InB}}^{i,t} \leq \overline{T}^{i,t}.
\end{equation}

\subsubsection{Power Supply}
The residential community $i$ can draw power from the grid, denoted as $p_{\text{grid}}^{i,t}$, and also from local PV generation $p_{\text{re,local}}^{i,t}$ and local energy market $p_{\text{ETB}}^{i,t}$. These power supplies satisfy the following constraints
\begin{align}\label{eq8}
    &0 \leq p_{\text{grid}}^{i,t} \leq \overline{P}_{\text{grid}}^{i},\\
    &0 \leq p_{\text{re,local}}^{i,t} + p_{\text{re,feed}}^{i,t} \leq \overline{P}_{\text{re}}^{i,t}, \\
    &0 \leq p_{\text{ETB}}^{i,t} \leq \overline{P}_{\text{ETB}}^{i},
\end{align}
where $\overline{P}_{\text{grid}}^{i}$ is the maximum power supply from the grid to community $i$ at all times; $\overline{P}_{\text{re}}^{i,t}$ is the local PV generation at $t$; $\overline{P}_{\text{ETB}}^{i,t}$ represents the maximum amount of purchase from the local market. Note that the PV generation can be used to serve the community, denoted as $p_{\text{re,local}}^{i,t}$ and also fed into the grid, denoted as $p_{\text{re,feed}}^{i,t}$. Given the feed-in tariff $\pi_{\text{re,feed}}^{t}$, the PV feed-in revenue is
$
\mathbf{R}_{\text{re}}^{i} = \sum_{t\in \mathcal{H}} \pi_{\text{re,feed}}^{t} p_{\text{re,feed}}^{i,t} \Delta t.
$

The power supply should meet the demand for each community $i$ at $t$ respectively, which leads to the power balance constraint as
\begin{equation}\label{eq9}
        p_{\text{\text{grid}}}^{i,t} + p_{\text{re,local}}^{i,t} + p_{\text{ETB}}^{i,t} + p_{\text{B2B}}^{i,t} = p_{\text{Batc}}^{i,t} + P_{\text{Bd,load}}^{i,t} + p_{\text{AC}}^{i,t},
\end{equation}
where the left-hand side includes the power supply from the grid, local PV, the local energy market, and battery supply power to building $p_{\text{B2B}}^{i,t}$. The right-hand side represents the total demand including the battery charging load, inflexible load in building $i$, and the central HVAC unit load. Note that when the battery performs discharge, it supplies power to local building first and then sells it in the local energy market $p_{\text{ETB}}^{i,t}$ or conduct battery-to-grid $p_{\text{B2G}}^{i,t}$.

We evaluate two types of electricity tariffs for the community as follows.
\begin{enumerate}
    \item Two-part tariff pricing (TPT) consists an energy price $\pi_\text{g}$ and a peak power price $\pi_{\text{peak}}$ for peak shaving. Hence, community $i$'s electricity cost is 
    $
    \mathbf{C}^{i}_{\text{grid}} = \pi_g\sum_{t\in \mathcal{H}} p_{\text{grid}}^{i,t} \Delta t + \pi_{\text{peak}}\max_{t\in \mathcal{H}} p_{\text{grid}}^{i,t}.
    $
    \item Time-of-use pricing (TOU) comprises TOU prices $\pi_g^t$ in peak, off-peak, and shoulder hours. Community $i$'s electricity cost is 
    $
    \mathbf{C}^{i}_{\text{grid}} = \sum_{t\in \mathcal{H}} \pi_g^t p_{\text{grid}}^{i,t} \Delta t.
    $
\end{enumerate}

\subsection{Power Network Model}
Communities are connected with each other on a distribution network. We consider a widely used linearized distribution network model \cite{zhong2018topology} to set network constraints for the active power, reactive power, and voltage as
\begin{align}
    & p^{i+1,t} = p^{i,t} - (p_{\text{grid}}^{i,t} + p_{\text{ETB}}^{i,t} - p_{\text{ETS}}^{i,t} - p_{\text{B2G}}^{i,t} - p_{\text{re,feed}}^{i,t})\label{eq10},\\
    &\underline{P}^{i}_{\text{DN}} \leq  p^{i,t} \leq \overline{P}^{i}_{\text{DN}}\label{eq11},\\
    & q^{i+1,t}  = q^{i,t} - Q_{\text{Load}}^{i,t}\label{eq12},\\
    & \underline{Q}^{i}_{\text{DN}} \leq  q^{i,t} \leq \overline{Q}^{i}_{\text{DN}}\label{eq13},\\
    & v^{i+1,t}  =  v^{i,t} - \left( R^{i+1} p^{i+1,t} + X^{i+1} q^{i+1,t} \right) / V^{0,t}\label{eq14},\\
    &\underline{V}^{i} \leq  v^{i,t} \leq \overline{V}^{i}\label{eq15},
\end{align}
where $p^{i,t}$ is the active power flow from node $i-1$ to $i$ in time slot $t$.
The maximum and minimum allowed active power flow from node $i-1$ to $i$ are $\overline{P}^{i}_{\text{DN}}$ and $\underline{P}^{i}_{\text{DN}}$, respectively. The reactive power flow is denoted as $q^{i,t}$ from node $i-1$ to $i$ at $t$, and $Q_{\text{Load}}^{i,t}$ is reactive load at node $i$ in $t$. The maximum and minimum allowed reactive power flow from node $i-1$ to $i$ are $\overline{Q}^{i}_{\text{DN}}$ and $\underline{Q}^{i}_{\text{DN}}$, respectively. The voltage of node $i$ at $t$ is denoted as $v^{i,t}$, $V^{0,t}$ is the voltage of node $0$ at $t$, $R^i$ and $X^i$ are the resistance and reactance of branch connected to node $i-1$ and $i$, respectively. The node voltage needs to be kept between $\overline{V}^{i}$ and $\underline{V}^{i}$, which are the lower bound and upper bound for the voltage, respectively.

\subsection{Value-Stacking Optimization Problem Formulation}\label{Optimal Problem Formulation}
The community battery can interact with the local building, grid, and local market for value stacking. We denote $p_{\text{B2B}}^{i,t}$, $p_{\text{B2G}}^{i,t}$, and $p_{\text{ETS}}^{i,t}$ as battery discharge to perform battery-to-building (B2B), battery-to-grid (B2G) and energy trading in the local market. The community battery discharging satisfies the following constraint
\begin{equation}\label{eq16}
    p_{\text{Batd}}^{i,t} = p_{\text{B2B}}^{i,t} + p_{\text{B2G}}^{i,t} + p_{\text{ETS}}^{i,t}.
\end{equation}

\begin{enumerate}
\item Value stream of battery-to-building:
The stored energy can be discharged to serve building load when the tariffs are high \cite{li2023review}. The B2B power $p_{\text{B2B}}^{i,t}$ is bounded by the maximum allowed supply from the battery to buildings, denoted as $\overline{P}_{\text{B2B}}^{i,t}$, i.e.,
\begin{equation}\label{eq17}
    0 \leq p_{\text{B2B}}^{i,t} \leq \overline{P}_{\text{B2B}}^{i,t}.
\end{equation}

\item Value stream of battery-to-grid: The battery energy can also be discharged to feed into the grid, bounded by the maximum battery-to-building power $\overline{P}_{\text{B2G}}^{i,t}$ as
\begin{equation}\label{eq18}
    0 \leq p_{\text{B2G}}^{i,t} \leq \overline{P}_{\text{B2G}}^{i,t}.
\end{equation}
Given the the dynamic price $\pi_{\text{B2G}}^t$ for the feed-in power, the B2G revenue is $\mathbf{R}^{i}_{\text{B2G}} = \sum_{t\in \mathcal{H}} \pi_{\text{B2G}}^{t} p_{\text{B2G}}^{i,t}\Delta t$.

\item Value stream of energy trading in local market:
Each community can exchange power with other communities using community batteries through the local market. Given the energy trading prices $\pi_{\text{ET}}^{t}$ set as a mid-market rate \cite{long2017peer}, community $i$'s net cost (defined as the cost deducted by the revenue) for energy trading is
$
\mathbf{C}_{\text{ET}}^{i} = \sum_{t\in \mathcal{H}} \pi_{\text{ET}}^{t} (p_{\text{ETB}}^{i,t} - p_{\text{ETS}}^{i,t}) \Delta t.
$
Note that each community $i$ cannot sell power to the local market and buy from it at the same time, we introduce binary variables $y^{i,t}\in \{ 0, 1\}$ to constrain energy exchange in 
\begin{align}
    &0 \leq p_{\text{ETS}}^{i,t} \leq \overline{P}_{\text{ETS}}^{i} y^{i,t}, \label{eq19}\\
    &0 \leq p_{\text{ETB}}^{i,t} \leq \overline{P}_{\text{ETB}}^{i}(1-y^{i,t}),\label{eq20}
\end{align}
where $\overline{P}_{\text{ETS}}^{i}$ and $\overline{P}_{\text{ETB}}^{i}$ represent the upper bounds for selling and buying energy in the local market for community $i$. 

The total energy selling and buying in the local energy market should satisfy the balance constraint in each time slot $t$ as 
\begin{equation}\label{eq21}
    \sum_{i \in \mathcal{I}} (p_{\text{ETS}}^{i,t} - p_{\text{ETB}}^{i,t}) = 0.
\end{equation}
\end{enumerate}

After presenting the system model, we obtain $\mathbf{C}_{S1}^{\text{total}}$ as the total cost of all communities to be minimized and formulate the community battery value-stacking optimization problem as 
\begin{equation}
\begin{aligned} \label{S1}
\min ~& \mathbf{C}_{S1}^{\text{total}} = \sum_{i \in \mathcal{I}}(\mathbf{C}^{i}_{\text{grid}}+\mathbf{C}_{\text{battery}}^{i}+\mathbf{C}_{\text{ET}}^{i}-\mathbf{R}_{\text{B2G}}^{i}-\mathbf{R}_{\text{re}}^{i})\\
\text{s.t.} ~& \eqref{eq1}-\eqref{eq21}.
\end{aligned}
\end{equation}

\section{Distributed Optimization Algorithm Design}\label{Disalo}
To solve the community battery value-stacking optimization problem, we develop an asynchronous distributed ADMM algorithm, which can work under asynchronous communications while maintaining the operational privacy of communities. The need for an asynchronous distributed optimization algorithm arises from the fact that different communities have varying communication capacities. The standard ADMM can facilitate distributed operation to preserve communities' privacy, but it relies on synchronous communications among all communities, which is not realistic or efficient. 

We define $\mathbf{p}_{\text{ex}}^{i} \triangleq [\mathbf{p}_{\text{B2G}}^{i},\mathbf{p}_{\text{ETS}}^{i},\mathbf{p}_{\text{ETB}}^{i} ,\mathbf{p}_{\text{grid}}^{i},\mathbf{p}_{\text{re,feed}}^{i}]^\mathsf{T}$ and introduce auxiliary variables $\widetilde {\mathbf{p}}_{\text{ex}}^{i} \triangleq [\widetilde{p}_{\text{ex}}^{i,1}, \widetilde{p}_{\text{ex}}^{i,2},\cdots, \widetilde{p}_{\text{ex}}^{i,H}]^\mathsf{T}$ and dual variables $\boldsymbol{\lambda}^{i} \triangleq [\lambda^{i,1}, \lambda^{i,2},\cdots, \lambda^{i,H}]^\mathsf{T}$ to derive the augmented Lagrangian function as
\begin{align}\label{eq:L}
    \mathcal{L} = &\sum_{i \in \mathcal{I}}(\mathbf{C}^{i}_{\text{grid}}+\mathbf{C}_{\text{battery}}^u+\mathbf{C}_{\text{ET}}^{i}-\mathbf{R}_{\text{B2G}}^{i} - \mathbf{R}_{\text{re}}^{i}) \notag \\ 
    & +\sum_{i\in \mathcal{I}}\left[(\boldsymbol{\lambda}^{i})^\mathsf{T} (\widetilde{ \mathbf{p}}_{\text{ex}}^{i} - \mathbf{p}_{\text{ex}}^{i}) + \frac{\rho}{2}\left\|\widetilde {\mathbf{p}}_{\text{ex}}^{i} - \mathbf{p}_{\text{ex}}^{i}\right\|_2 ^2 \right],
\end{align}
where $\rho > 0$ is the penalty coefficient.

Based on the augmented Lagrangian \eqref{eq:L}, we decompose the community battery value-stacking optimization problem into sub-problems for each community and the grid operator.
In communities' sub-problems, each community minimizes its cost in parallel by taking the auxiliary variable $\widetilde {\mathbf{p}}_{\text{ex}}^{i}$ and dual variable $\boldsymbol{\lambda}^{i}$ as given. Each community's sub-problem is formulated as
\begin{align}
    &\mathop{\min}\  (\mathbf{C}^{i}_{\text{grid}}+\mathbf{C}_{\text{battery}}^{i}+\mathbf{C}_{\text{ET}}^{i}-\mathbf{R}_{\text{B2G}}^{i} - \mathbf{R}_{\text{re}}^{i}) \notag \\
     & + \left[(-\boldsymbol{\lambda}^{i})^\mathsf{T} \mathbf{p}_{\text{ex}}^{i} + \frac{\rho}{2}\left\| \widetilde {\mathbf{p}}_{\text{ex}}^{i} - \mathbf{p}_{\text{ex}}^{i}\right\|_2^2 \right] \notag \\
     &\text{s.t.}~ (\ref{eq1})-(\ref{eq9}),\text{and}~(\ref{eq16})-(\ref{eq20}) \notag \\
     & \text{variables}:~ \mathbf{p}_{\text{grid}}^u, \mathbf{p}_{\text{re,local}}^{i}, \mathbf{p}_{\text{re,feed}}^{i}, \mathbf{p}_{\text{Batc}}^{i},\mathbf{p}_{\text{Batd}}^{i},\mathbf{p}_{\text{B2B}}^{i},\mathbf{p}_{\text{B2G}}^{i},\mathbf{p}_{\text{ETS}}^{i},\notag \\
     &\mathbf{p}_{\text{ETB}}^{i},\mathbf{p}_{\text{ex}}^{i},\mathbf{p}_{\text{AC}}^{i}, \mathbf{T}_{\text{InB}}^{i}, \mathbf{x}^{i},\mathbf{y}^{i}.
\end{align}

Each community $i$ will solve the primal variables $\mathbf{p}_{\text{ex}}^{i}$ and submit them to the grid operator for updating the dual variables $\boldsymbol{\lambda}^{i}$ and the auxiliary variables $\widetilde{\mathbf{p}}_{\text{ex}}^{i}$.

The grid operator will manage the power network constraint through solving the grid operator's sub-problem and update the dual variables $\boldsymbol{\lambda}^{i}$ and the auxiliary variables $\widetilde{\mathbf{p}}_{\text{ex}}^{i}$. The operator obtains primal variables $\mathbf{p}_{\text{ex}}^{i}$ from community $i \in \mathcal{I}$ and then calculates the auxiliary variables $\widetilde{\mathbf{p}}_{\text{ex}}^{i}$ by solving 
\begin{align}\label{eq26}
    &\mathop{\min}\sum_{i\in \mathcal{I}}\left[(\boldsymbol{\lambda}^{i})^\mathsf{T} \widetilde{ \mathbf{p}}_{\text{ex}}^{i} + \frac{\rho}{2}\left(\widetilde {\mathbf{p}}_{\text{ex}}^{i} - \mathbf{p}_{\text{ex}}^{i}\right)^2 \right] \notag \\
    &\text{s.t.}~(\ref{eq10})-(\ref{eq15}),\text{and}~(\ref{eq21}) \notag \\
    &\text{variables}:~ \{ \widetilde{\mathbf{p}}_{\text{ex}}^{i},\mathbf{p}^{i},\mathbf{q}^{i},\mathbf{v}^{i}, \forall i \in \mathcal{I} \}.
\end{align}

Different from the standard ADMM, we propose an asynchronous distributed ADMM algorithm \cite{zhang2014asynchronous}. Different from the standard ADMM algorithm requiring synchronous communications for all communities and the grid operator to complete the information exchange in each iteration, our asynchronous distributed ADMM algorithm allows asynchronous communications, being more flexible and robust against latency in information exchange between communities and the grid operator. When the grid operator computes the dual and auxiliary variables, it does not have to wait for updates from all communities before making updates and broadcasting to communities. Those communities failing to exchange information with the grid in one iteration can update their decisions to the operator in future iterations, allowing for a more flexible and robust information-sharing paradigm. 

We denote $\mathcal{I}(k)$ as the set of communities which successfully receive the corresponding auxiliary variables $\widetilde{\mathbf{p}}_{\text{ex}}^{i}$ and dual variables $\boldsymbol{\lambda}^{i}$ and could update the primal variables $\mathbf{p}_{\text{ex}}^{i}$ in iteration $k$. For those communities which fail to do so, the grid operator can use these communities' previous updates when the required information exchange is missing. As Eq. (\ref{eq26}) presents, community $i$'s updates $\mathbf{p}_{\text{ex}}^{i}$ in iteration $k$ are obtained as
\begin{equation}\label{eq27}
    \mathbf{p}_{\text{ex}}^{i}(k+1) = \left \{
    \begin{aligned}
        & \arg\min\mathcal{L}(\mathbf{p}_{\text{ex}}^{i},\widetilde{\mathbf{p}}_{\text{ex}}^{i},\boldsymbol{\lambda}^{i,t}),~\text{if}~ i\in \mathcal{I}(k), \\
        & \mathbf{p}_{\text{ex}}^{i}(k)~,~ \text{if}~ u\notin \mathcal{I}(k), 
    \end{aligned}
   \right.
\end{equation}
which accommodate asynchronous communications between communities and the grid, showing better reliability and robustness in the presence of latency. Last, the grid operator updates the dual variables as follows
\begin{align}\label{eq28}
    \boldsymbol{\lambda}^{i}(k+1):= \boldsymbol{\lambda}^{i}(k) + \rho\left(\widetilde {\mathbf{p}}_{\text{ex}}^{i}(k) - \mathbf{p}_{\text{ex}}^{i}(k)\right),
\end{align}
and broadcasts it to all communities in each iteration.

\begin{figure}[t]
\centering
\epsfig{figure=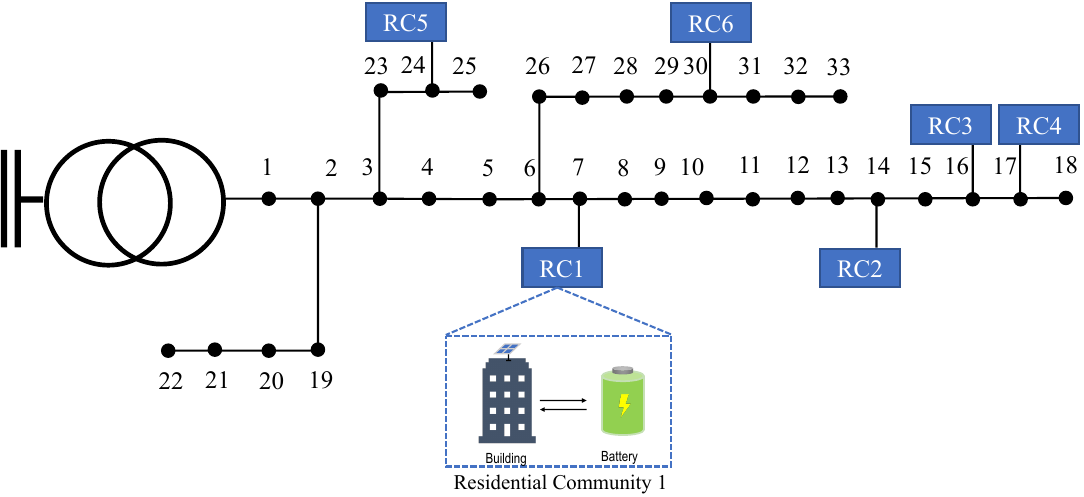,width=0.6\linewidth} 
\vspace{-2mm}
\caption{IEEE 33 bus radial distribution test system.}
\label{fig:IEEE33BUS}
\vspace{-2mm}
\end{figure}

\section{Simulation and Discussion}\label{Simulation and Discussions}
We validate our model and algorithm on the IEEE 33 bus test system, as depicted in Fig. \ref{fig:IEEE33BUS}, where six communities connect to nodes 7, 14, 16, 17, 24, and 30, respectively. In each community, the community battery capacity is $500$kWh, the initial stored energy ranges from $230$kWh to $270$kWh following a truncated normal distribution, and the maximum charging power is $100$kW. We use real market data from the National Energy Market (NEM) in Australia and evaluate two tariffs, e.g., TOU and TPT. 

\subsection{Convergence and Optimality of Asynchronous ADMM}\label{Convergence of A-ADMM}
\begin{figure}[t]
\centering
\epsfig{figure=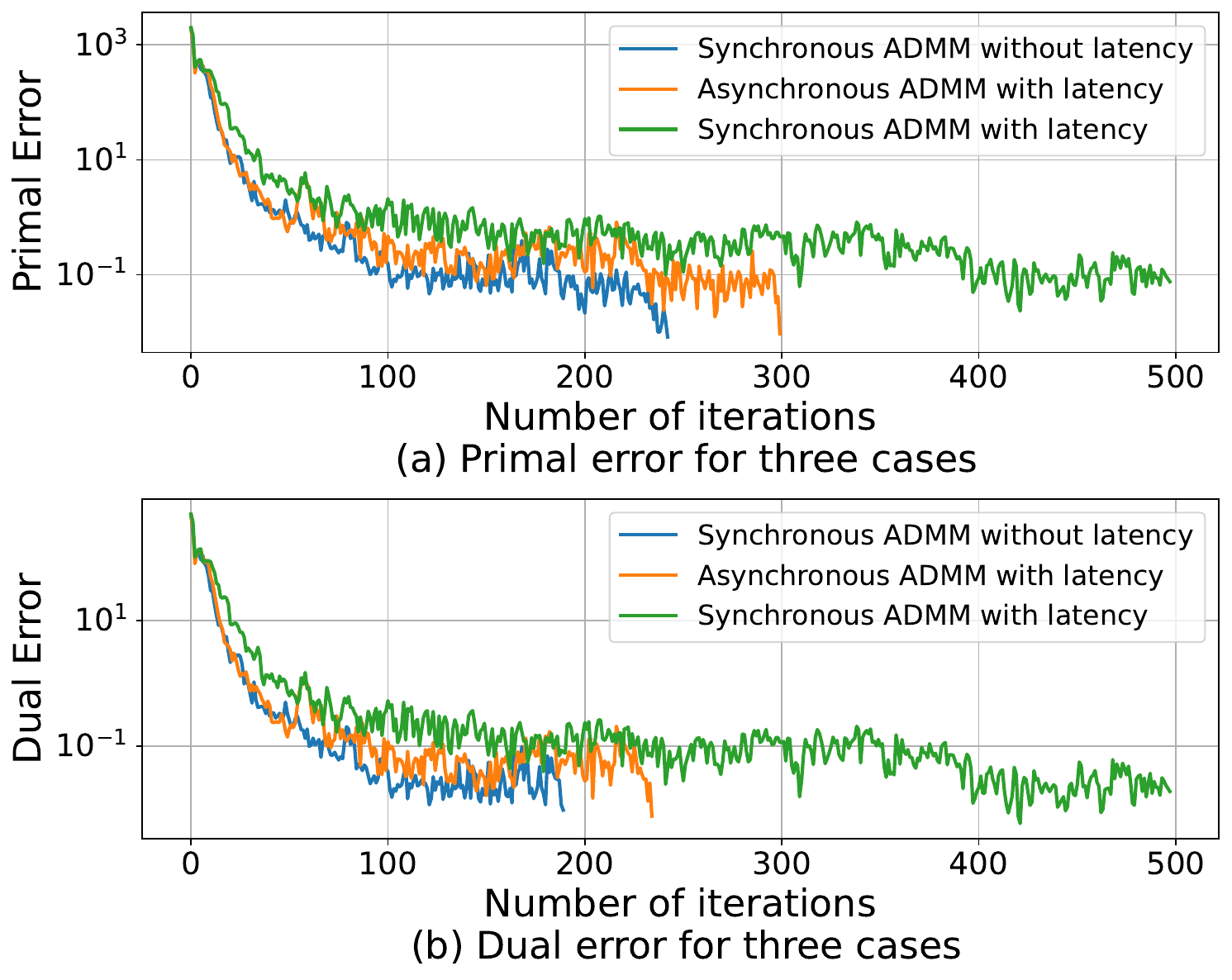,width=0.6\linewidth} 
\vspace{-2mm}
\caption{Comparison of the synchronous and asynchronous ADMM algorithm.}
\label{fig:Comparison of three ADMM}
\end{figure}

Firstly, we illustrate the convergence of our proposed asynchronous ADMM algorithm with communication latency, compared to synchronous ADMM without and with communication latency. For synchronous ADMM with communication latency, different from Eq.~(\ref{eq27}), when there is a delay in communicating decision variables $\mathbf{p}_{\text{ex}}^{i}$, the grid operator sets 0 for $\mathbf{p}_{\text{ex}}^{i}$ in iteration $k$. We set the convergence thresholds as $\epsilon_1=0.01$ and $\epsilon_2=0.01$, and the algorithm terminates when both thresholds are reached. As Fig~\ref{fig:Comparison of three ADMM} shows, we see the synchronous ADMM without communication latency converges fast. But it is not practical as real-world environment always has communication latency. The synchronous ADMM fails to converge community even after 500 iterations when there is a communication latency. In contrast, our proposed asynchronous ADMM converges after 300 iterations in realistic scenarios with communication latency.

Second, we assess the optimally of the solution obtained from the proposed algorithm by comparing it to the optimal solution solved by a centralized method. The centralized method solves the problem in (\ref{S1}) using the Gurobi solver and attains the optimal solution serving as a benchmark. When the convergence thresholds $\epsilon_1$ and $\epsilon_2$ are set to $0.01$, the deviation from the benchmark is $0.33\%$. Setting $\epsilon_1$ = $\epsilon_2$ = $0.001$ reduces the deviation in the cost to $0.067\%$, resulting in slightly better performance. However, the algorithm requires $993$ iterations to converge. For a good balance between cost minimization and computational efficiency, we choose the thresholds $\epsilon_1$ and $\epsilon_2$ as $0.01$ for the remaining simulations.

\subsection{Performance of Battery Value Stacking}
\begin{figure}[!t]
  \begin{minipage}[t]{0.5\linewidth}
    \centering
    \includegraphics[scale=0.3]{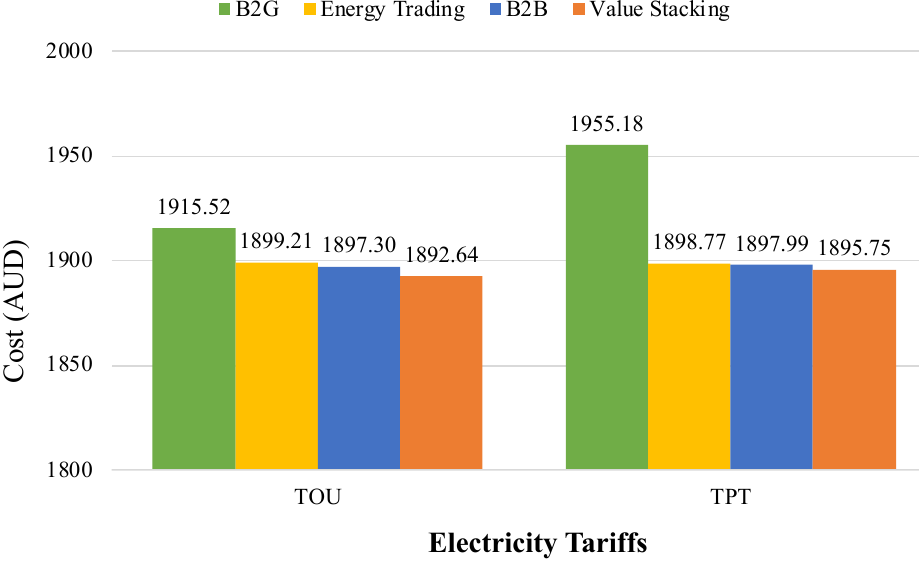}
    \caption{Communities' cost under\\ two tariffs in four scenarios.}
    \label{fig:label3}
  \end{minipage}%
  \begin{minipage}[t]{0.5\linewidth}
    \centering
    \includegraphics[scale=0.3]{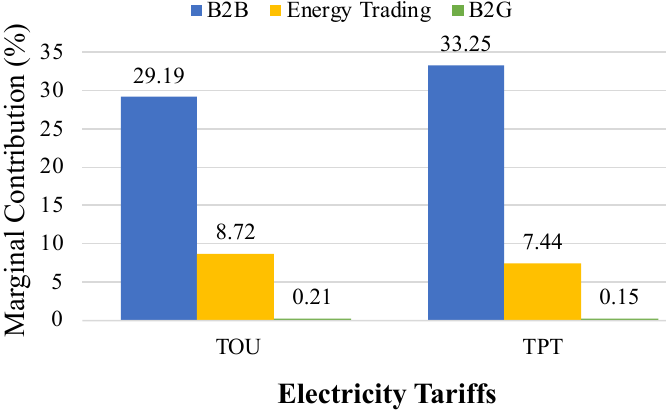}
    \caption{Marginal contribution of different value streams in two tariffs.}
    \label{fig:label4}
  \end{minipage}
  \vspace{-5mm}
\end{figure}

The performance of value stacking of community battery is presented in Fig. \ref{fig:label3}, where the vertical axis represents the total cost (AUD) for value stacking and three baselines. The two clusters of bars on the horizontal axis correspond to TOU and TPT. Under TOU pricing, value stacking achieves the lowest cost at $1892.64$ AUD, while B2G incurs the highest cost at $1915.52$ AUD. For TPT, value stacking registers a cost of $1895.75$ AUD, slightly higher compared to TOU. Energy trading shows a reduced cost advantage under TPT in comparison to TOU, whereas B2B performs better under TOU than TPT. The TPT tariff aims to encourage peak shaving, resulting in a more evenly distributed energy purchase from the main grid to circumvent peak prices, differing from the characteristics of the TOU tariff.

To assess the contribution of each value stream within our value-stacking problem, we compute their marginal contributions under two tariffs, TOU and TPT. The marginal contribution is determined by the relative difference in cost reductions between the value stacking and baseline problems when excluding a single value stream. As depicted in Fig. \ref{fig:label4}, the marginal contribution of B2B significantly surpasses that of both B2G and energy trading, reaching levels of up to $29.19\%$ and $33.25\%$, respectively, under both TOU and TPT tariffs. Additionally, communities prioritize selling power to the local market during peak electricity consumption hours over supplying power to B2G, resulting in a relatively minor contribution of B2G under both tariffs. Therefore, B2B offers considerable benefits to communities under both TOU and TPT tariffs compared to energy trading and B2G.

\section{Conclusion and Future Work}\label{Conclusion and Future Work}
In this paper, we proposed a community battery value-stacking optimization problem considering local network constraints to maximize the value of battery storage systems in residential communities. In addition, we design a distributed optimization algorithm based on the asynchronous ADMM to solve the community battery value-stacking optimization problem, which can converge in realistic scenarios with asynchronous way to exchange information. The results showed the value-stacking model achieved better performance under both TPT and TOU.

\bibliographystyle{IEEEtran}
\bibliography{ref.bib}

\end{document}